\documentclass[dvips]{acta}
\usepackage{lscape,epsfig}
\usepackage{amssymb}
\usepackage{amsmath}

\SetPages{0}{0}

\SetVol{66}{2016}

\begin{document}

\begin{Titlepage}
\Title{OGLE Study of the Sagittarius Dwarf Spheroidal Galaxy\\ 
and its M54 Globular Cluster
\footnote{Based on observations obtained with the 1.3-m Warsaw telescope
at the Las Campanas Observatory of the Carnegie Institution for
Science.}}

\Author{A.~~H~a~m~a~n~o~w~i~c~z$^1$,~~P.~~P~i~e~t~r~u~k~o~w~i~c~z$^1$,~~A.~~U~d~a~l~s~k~i$^1$,\\
P.~~M~r~\'o~z$^1$, I.~~S~o~s~z~y~\'n~s~k~i$^1$,~~M.~K.~~S~z~y~m~a~\'n~s~k~i$^1$,\\
J.~~S~k~o~w~r~o~n$^1$,~~R.~~P~o~l~e~s~k~i$^{1,2}$,~~\L{}.~~W~y~r~z~y~k~o~w~s~k~i$^1$,\\
S.~~K~o~z~\l{}~o~w~s~k~i$^1$,~~M.~~P~a~w~l~a~k$^1$~~and~~K.~~U~l~a~c~z~y~k$^3$}
{$^1$ Warsaw University Observatory, Al. Ujazdowskie 4, 00-478 Warszawa, Poland\\
e-mail:(ahamanowicz,pietruk,udalski)@astrouw.edu.pl\\
$^2$ Department of Astronomy, Ohio State University, 140 W. 18th Ave., Columbus, OH 43210, USA\\
$^3$ Department of Physics, University of Warwick, Gibbet Hill Road, Coventry, CV4 7AL, UK}

\Received{May 11, 2016}
\end{Titlepage}

\Abstract{We use the fundamental-mode RR Lyr-type variable stars (RRab)
from OGLE-IV survey to draw a 3D picture of the central part of the tidally
disrupted Sagittarius Dwarf Spheroidal (Sgr dSph) galaxy. We estimate
the line-of-sight thickness of the Sgr dSph stream to be
FWHM$_{\rm cen}=2.42$ kpc. Based on OGLE-IV observations collected
in seasons 2011--2014 we conduct a comprehensive study of stellar
variability in the field of the globular cluster M54 (NGC~6715) residing
in the core of this dwarf galaxy. Among the total number of 268
detected variable stars we report the identification of 174 RR~Lyr
stars, four Type II Cepheids, 51 semi-regular variable red giants, three
SX Phe-type stars, 18 eclipsing binary systems. Eighty-three variable stars
are new discoveries. The distance to the cluster determined from RRab stars
is $d_{\rm M54}=26.7 \pm0.03_{\rm stat} \pm1.3_{\rm sys}$ kpc.
From the location of RRab stars in the period--amplitude (Bailey) diagram
we confirm the presence of two old populations, both in the cluster
and the Sgr dSph stream.}

{\textit{Galaxies: individual: Sgr dSph -- globular clusters:
individual: M54 (NGC~6715) -- Stars: variables: RR Lyr, Type II Cepheids}}

\section{Introduction}
The Sagittarius Dwarf Spheroidal (Sgr dSph) galaxy, discovered by Ibata
\etal (1994), is a tidally disrupted stellar system colliding with the
Milky Way. Streams of this dwarf galaxy are extended around in the sky
(Majewski \etal 2003, Belokurov \etal 2006, Torrealba \etal 2015).
Studies of the spatial location of the streams help to test models of
the dark matter distribution in our Galaxy (Belokurov \etal 2014).

Globular cluster M54 is considered to be the actual core of the
Sgr dSph galaxy. It is located at the Galactic coordinates
$(l,b)=(+5\zdot\arcd607,-14\zdot\arcd087)$. However, the question where
this cluster was formed remains open. Photometric observations conducted
in 1990s by Sarajedini and Layden (1995) and also Layden and Sarajedini
(1997) indicated that M54 is physically associated with the Sgr dwarf.
Later studies by Majewski \etal (2003) and Monaco \etal (2005)
demonstrated that there is a nuclear condensation in the Sgr system
independent of the presence of the metal-poor population identified
with the globular cluster M54. Bellazzini \etal (2008) compared
radial velocities of 1152 candidate red giant branch stars from
Sgr dSph and M54, obtaining significantly different velocity dispersion
profiles in both systems. They concluded that the actual Sgr dSph
nucleus was most likely formed independently of M54. Searches for
variable stars, particularly RR Lyr-type stars being tracers of old
populations, may help in the determination of physical parameters and
better understanding of the formation process of M54 and the whole Sgr
dSph stream (\eg Kunder and Chaboyer 2009, Zinn \etal 2014).

First wide-field searches for RR Lyr stars in the main part of the Sgr
dSph galaxy were undertaken by the Optical Gravitational Lensing
Experiment (OGLE) and Disk Unseen Objects (DUO) surveys, already in the
year in which the dwarf galaxy was discovered. Mateo \etal (1995ab) reported
seven RR Lyr stars from the OGLE-I survey, very likely members of the Sgr
dSph, and estimated the distance to the dwarf of $25.2\pm2.8$ kpc
assuming a RR Lyr absolute brightness of $M_V=+0.6$~mag. Later,
Cseresnjes \etal (2000) presented the discovery of about 1500
fundamental-mode (RRab) pulsators in a 50 deg$^2$ area of the Sgr dwarf
monitored by DUO. Based on averaged periods obtained for RRab and also
RRc (first-overtone) pulsators Cseresnjes (2001) placed the Sgr dSph
galaxy in the long-period tail of the Oosterhoff I (OoI) group. From
the period and amplitude distributions of the RR Lyr population they found
an average metallicity of [Fe/H] $=-1.6$ dex with a contribution of a
minor population with [Fe/H] below $-2$ dex. At the same time,
metallicity gradients were noticed by Alard (2001) who compared red
giant branches in two fields separated by $6\arcd$ along the Sgr dSph
stream. Similar findings were announced by Zinn \etal (2014) who
surveyed about 840 deg$^2$ of the Galactic halo including the Sgr dSph
stream: RRab stars belonging to the stream form a mixture of OoI and
OoII groups with heavy weight toward OoI. From 208 RRab variable stars
observed by the MACHO survey (Alcock \etal 1997) and associated with the
Sgr dSph galaxy Kunder and Chaboyer (2009) derived the distance to its
northern extension. They obtained $24.8\pm0.8$ kpc and concluded that
the extension of the Sgr dSph galaxy toward the Galactic plane is
inclined toward us.

Recently, Soszy\'nski \etal (2014) reported the detection of 2286 RR Lyr
variable stars in a 9.8 deg$^2$ area toward the main body of the Sgr dwarf
surveyed by the OGLE-IV project (Udalski \etal 2015). Among the variable stars
there are 1767 RRab, 499 RRc, and 20 RRd (double-mode) type stars.

First variable stars in the field of M54 globular cluster were discovered by Rosino
(1952) who photographically found 28 such objects. Among them there were
two Type II Cepheids and 15 candidate RR Lyr-type stars. Later, Rosino
and Nobili (1958) reported on another 54 variable stars, almost all of them
of RR Lyr type. A search for variable stars conducted by Layden and
Sarajedini (2000) provided 35 new objects, roughly half of which are RR
Lyr stars and the other half are variable red giants. Other, more recent
studies include Sollima \etal (2010) work who announced the discovery
of 94 new variable stars and
Montiel and Mighell (2010) reanalysis of archival Hubble Space Telescope
(HST) Wide Field Planetary Camera~2 (WFPC2) images of the center of the
cluster. They reported 50 candidates for RR Lyr stars. The HST observations
were obtained in 1999 and consist of six images in the F814W filter
(close to the standard \textit{I} passband) and six images in the F555W
filter (\textit{V} passband) spanning 7.5 hours total. Fourteen of the
proposed candidates can be assigned to the previously known variable stars. The
whole list of variable stars in the field of M54 is presented in Clement
\etal (2001) who made an on-line update in
2014\footnote{http://www.astro.utoronto.ca/{\textasciitilde{}}cclement/read.html}.
Three-fourths of all variable stars in this list are RR Lyr stars.

In recent years, two independent distance indicators were used to
measure the distance to the globular cluster M54: RR Lyr variable stars and
tip of the red giant branch (TRGB) stars. The following distance
estimates can be found in the literature: $27.4\pm1.5$ kpc (Layden and
Sarajedini 2000, based on RR Lyr stars), $26.30\pm1.8$ kpc (Monaco \etal
2004, TRGB stars), $26.7\pm1.1$ kpc (Sollima \etal 2010, RR Lyr stars).
According to the 2010 update of the Harris (1996) catalog, the center of
M54 is located at ($\alpha$,$\delta$)$_{2000.0}$=$(18\uph55\upm03\zdot\ups33$,
$-30\arcd28\arcm47\zdot\arcs5)$, the globular cluster has the core
radius $r_{\rm c}=0\zdot\arcm09$ and half-light radius $r_{\rm
h}=0\zdot\arcm82$. The cluster tidal radius $r_{\rm t}$ is estimated to
be about $7\zdot\arcm5$, according to the first edition of this catalog.

In this work, based on all detected RRab stars in the OGLE-IV Galactic
bulge fields we draw a 3D picture of the Sgr dSph galaxy in the
background of the old Milky Way bulge (Pietrukowicz \etal 2015). We
estimate the line-of-sight thickness of the Sgr stream. We also search
for any type of variable objects within the tidal radius of M54 using
observations spanning four-years from the OGLE-IV survey and by applying
non-standard reduction techniques. Based on RRab stars we determine the
distance to M54. Finally, using period--amplitude diagram, we compare
distributions of this type of stars residing in the core cluster with
properties of RR Lyr stars from the Sgr dSph galaxy.

It is worth noting that results of an independent search for
variable objects in the very central part of M54 (Figuera Jaimes \etal 2016)
have recently been announced almost simultaneously with the presented
here discoveries.

\section{Observations}
The OGLE project is the worldwide largest stellar variability sky
survey, with one trillion ($10^{12}$) single brightness measurements
collected and nearly half a million genuine variable stars discovered by
2016. The survey is conducted at Las Campanas Observatory, Chile, since
1992. OGLE monitors about 1.3 billion stars in the sky toward the Milky
Way bulge and disk and the Magellanic System in searches for periodic as
well as transient objects (\eg Soszy\'nski \etal 2013, 2014, 2015,
Pietrukowicz \etal 2013, Mr\'oz \etal 2015). For over 20 years OGLE
operates the dedicated 1.3-m Warsaw telescope. Since 2010 the project is
in its fourth phase, OGLE-IV. The OGLE-IV camera is a mosaic of 32
2K$\times$4K CCDs giving a total field of view of 1.4 deg$^2$ at the
scale of 0.26~arcsec/pixel. Most of images (about 90\%) are taken in the
$I$-band filter, while the remaining images are taken in the $V$-band
filter. Detailed description of the OGLE-IV survey, instrumentation and
data reduction can be found in Udalski \etal (2015).

During 2011--2014, OGLE observed seven fields (BLG705--BLG711) toward
the main part of the Sgr dSph galaxy. For each of the seven fields about
160 $I$-band and 10 $V$-band frames were collected, all with exposure
times of 150~s. Most of the $I$-band data were obtained in 2011. The
distribution of the OGLE-IV Galactic bulge fields with detected RR Lyr
stars are presented in Soszy\'nski \etal (2014). These observations
constitute a pilot study of the part of Sgr dSph stream crossing the
Milky Way. Currently, the OGLE project conducts a variability survey of
the outer Galactic bulge, including a much larger area of the Sgr dSph
stream (see Fig.~3 in Pietrukowicz 2016).

\section{3D Picture of the Sgr dSph Galaxy}

\subsection{Distance Estimation from RR Lyr stars}

RR Lyr variable stars are very useful distance indicators. Particularly, RRab
stars have several advantages over other types of these pulsators. In
comparison to RRc (and also very rare RRd) stars, RRab variable stars are
more numerous, on average are intrinsically brighter in $I$, have higher
amplitudes, and have characteristic saw-tooth-shaped light curves,
making the searches highly complete. There is one more very practical
photometric property of RRab stars: based on the pulsation period $P$
and a combination of Fourier decomposition phases
$\phi_{31}=\phi_3-3\phi_1$ of a star one can estimate its metallicity
[Fe/H] (Jurcsik 1995, Jurcsik and Kov\'acs 1996). Recently, RRab stars
from OGLE-IV have been used by Pietrukowicz \etal (2015) to study
three-dimensional structure of the old Galactic bulge. Here, we apply a
similar approach to determine distances to the RRab variable stars laying
between the half-light and tidal radii of the globular cluster M54 and
other RR Lyr variable stars in the Sgr dSph area observed by OGLE-IV (fields
BLG705--BLG711). We use the following sequence of equations to derive
physical properties and distance $d$ to the individual RRab stars:
\begin{eqnarray}
{\rm [Fe/H]} &=& -3.142 - 4.902~P + 0.824~\phi_{31} \\
{\rm log}~Z &=& {\rm [Fe/H]} - 1.765 \\
M_I &=& 0.471 - 1.132~{\rm log}~P + 0.205~{\rm log}~Z \\
M_V &=& 2.288 + 0.882~{\rm log}~Z + 0.108~({\rm log}~Z)^2 \\
(V-I)_0 &=& M_V - M_I \\
E(V-I) &=& 1.375~E(B-V) \\
A_I &=& 1.969~E(B-V) \\
I_0 &=& I - A_{I} \\
d &=& 10^{1+0.2(I_0-M_I)}
\end{eqnarray}
We use the empirical relation on metallicity (Eq. 1) derived for
$I$-band light curves by Smolec (2005). Relations on the absolute
brightnesses $M_I$ and $M_V$ (Eq. 3--4 together with Eq. 2) come from
the theoretical work by Catelan \etal (2004). Reddening values
$E(B-V)$ were taken from the extinction map in Schlegel \etal (1998),
corrected by a factor of 0.86 as suggested by Schlafly and Finkbeiner
(2011), and transformed to $E(V-I)$ and $A_I$ using extinction-law
coefficients from Schlegel \etal (1998). In Fig. 1, we present observed
and dereddened CMDs for about 22~900 OGLE RRab variable stars with measured
brightnesses in $V$ and $I$. In Fig. 2, we show histogram of the
obtained distances to these stars. Fig. 3 presents the 3D distribution
of all RRab variable stars. The origin of the coordinate system is in the Sun
and the $X$ axis points toward the Galactic center. Variable stars from
the Sgr dSph galaxy form a prominent group behind the Galactic bulge,
but they are mixed with Galactic halo and thick disk stars.

\begin{figure}[htb]
\includegraphics[width=12.5cm]{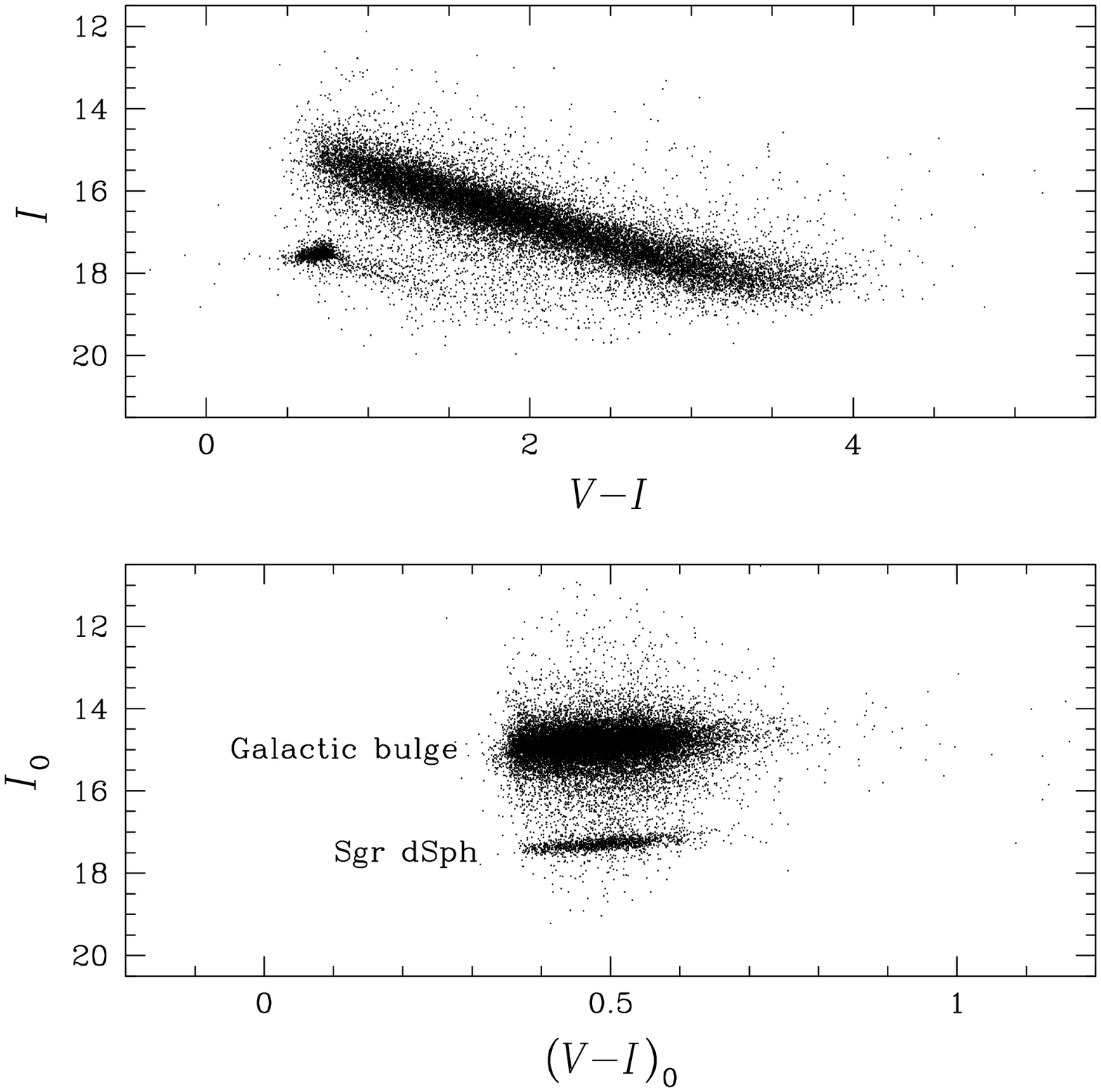}
\FigCap{Observed (upper panel) and dereddened (lower panel) CMDs for
RRab stars in the OGLE-IV bulge fields. Sgr dSph stars clearly separate
from the Galactic bulge variable stars.}
\end{figure}

\begin{figure}[htb]
\includegraphics[width=12.5cm]{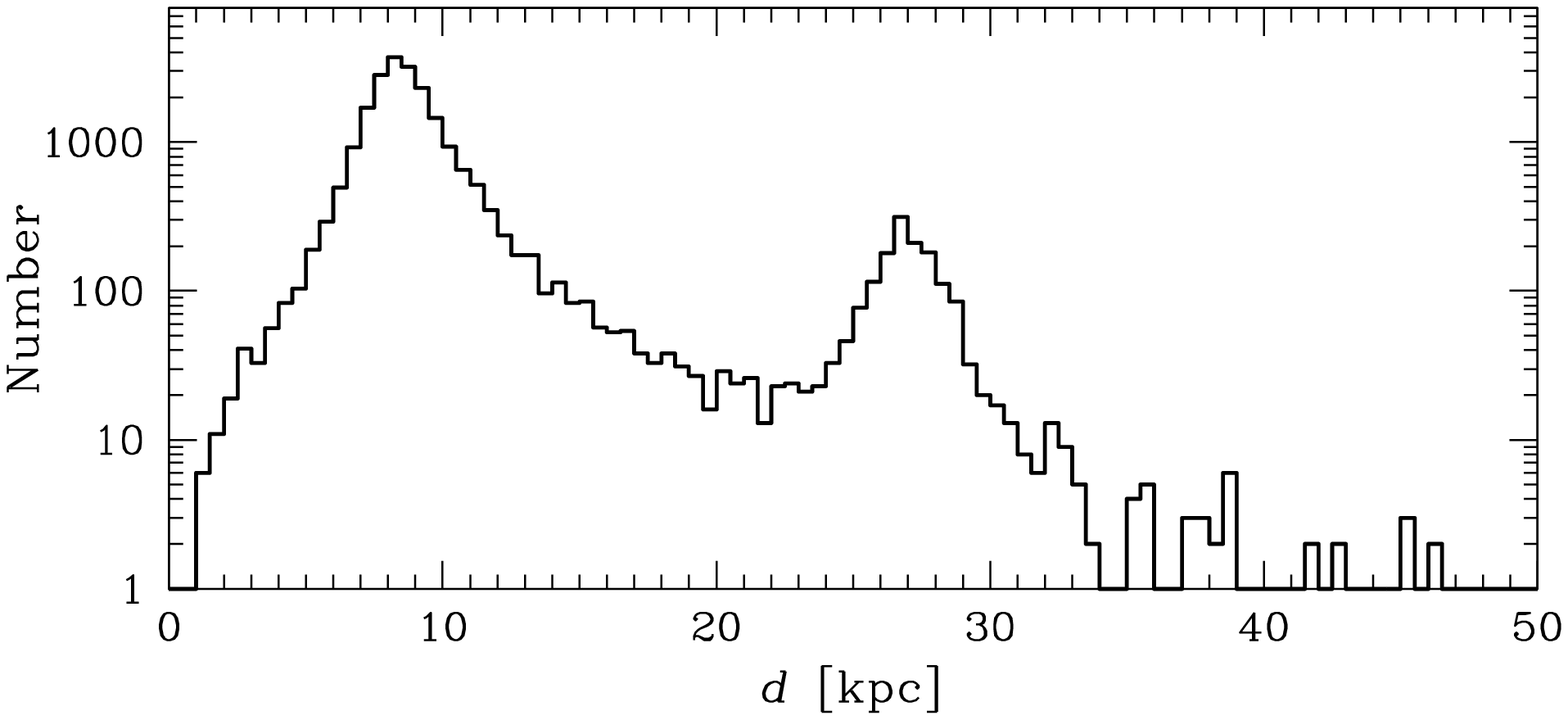}
\FigCap{Distance distribution for about 22~900 RRab stars observed in
the OGLE-IV Galactic bulge fields. The maxima correspond to the Galactic
bulge and roughly three times more distant Sgr dSph galaxy. The sharp
peak for the Sgr dwarf is mainly formed by the variable stars from the
globular cluster M54.}
\end{figure}

\begin{figure}[htb]
\includegraphics[width=12.5cm]{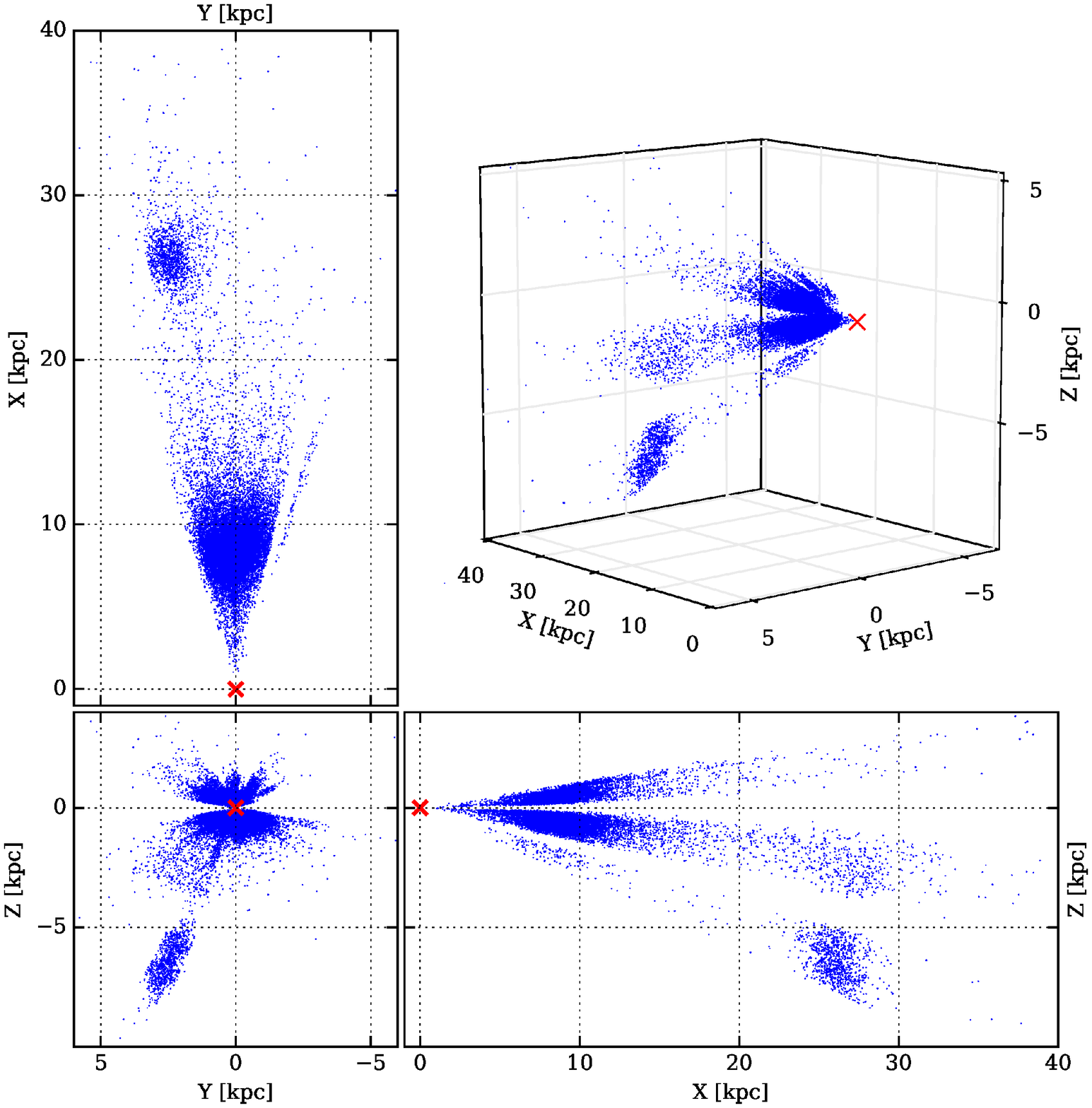} \\
\FigCap{Distribution of about 22~900 OGLE RRab stars in 3D space with
the origin in the Sun (marked with red cross). Most of the stars are
Galactic bulge objects. The second prominent structure is the Sgr dSph
galaxy.}
\end{figure}

\subsection{Thickness of the Sgr dSph Stream}

In Fig.~4, we present the results of estimating thickness of the Sgr
dSph stream in two observed directions: toward its core section and
behind the Galactic bulge. What we determine is the full width at
half maximum FWHM$\approx2.35\sigma$ of a Gaussian profile fitted to the
histogram of derived in Section 3.1 RR~Lyr distances. In the central
section the observed FWHM is 2.49~kpc. Variable stars from the globular
cluster M54 were not taken into account for this measurement. The part
of the Sgr dwarf behind the Galactic bulge is clearly superimposed
on the declining distribution of the Galactic halo and thick disk
RR Lyr stars. In this direction, the stream seems to be thicker with
the observed FWHM of 3.50~kpc (after correction for Galactic
background; \cf Fig.~4 -- middle panel). To obtain the true FWHM
of the Sgr stream, from the observed value we need to quadratically
subtract FWHM corresponding to the statistical uncertainty of the
distance, $2.35\sigma_{\rm d}$. If we adopt a mean accuracy of OGLE
brightness measurements of $\sigma_I=0.02$~mag, for the Sgr central
section we find an uncertainty of the distance $\sigma_{\rm d}\approx0.25$~kpc,
hence FWHM$_{\rm cen}\approx2.42$~kpc.
For the section behind the bulge, for which we use the $E(J-K_s)$
map from Gonzalez \etal (2012) with statistical $\sigma_{\rm E(J-Ks)}\approx0.01$~mag,
we find an uncertainty of the distance modulus of 0.040~mag ($\sigma_d \approx 0.5$~kpc)
and the true FWHM$_{\rm offcen}=3.30$~kpc.

Comparison of the distances in the two directions do not favor
the result from Kunder and Chaboyer (2009) that the Sgr stream
is inclined toward us closer to the Galactic plane.
However, one has to remember that our sample of Sgr dwarf
variable stars at lower Galactic latitudes is incomplete due to high
reddening close to the Galactic plane. Moreover, we applied different
dereddening procedures for objects located at $b<-10\arcd$ (maps from
Schlegel \etal 1998) and $b>-10\arcd$ (relation on extinction from Nataf
\etal 2013).

\begin{figure}[htb]
\includegraphics[width=12.5cm]{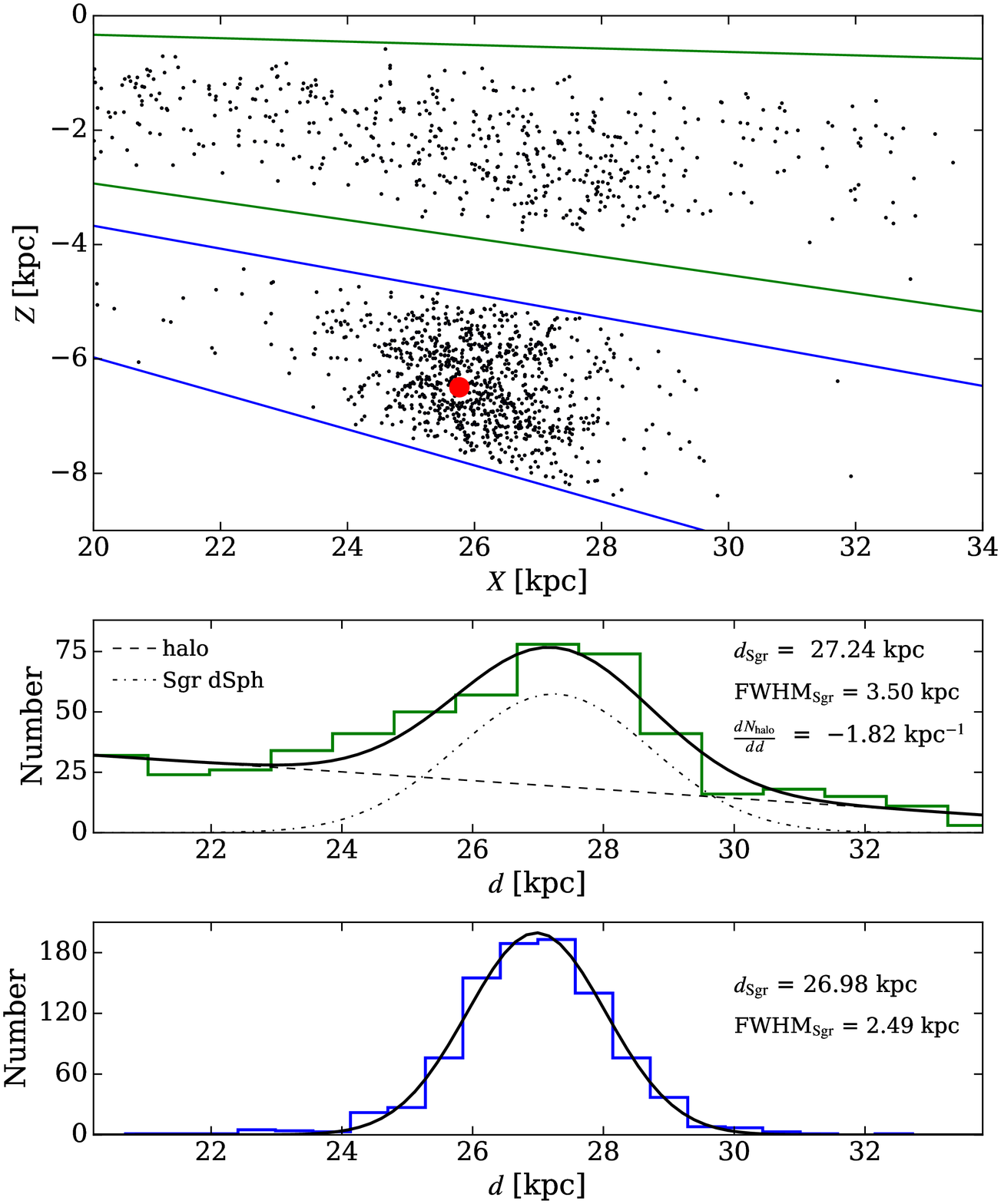}
\FigCap{Distance and observed line-of-sight thickness of the Sgr dSph
stream. The \textit{upper panel} shows RRab stars in the projection onto the $XZ$ plane.
The Sun and Galactic Center (GC) are on the left-hand side of \textit{this panel}.
Red circle marks the exact position of the globular cluster M54 as found
in Section~6. Histograms in the \textit{middle panel }and \textit{lower panel} show the
distance distributions in two directions delimited by green and blue
lines, respectively.}
\end{figure}

\section{Search for Variable Stars in the Field of M54}

The center of the globular cluster M54 can be found in the northern part
of the CCD \#09 in the OGLE field BLG708. The cluster half-light area is
well within this CCD, but the whole area delimited by the tidal radius
partially spans also over CCDs \#08, 10, 17, 18, and 19 in the same
field. Location of the OGLE CCDs together with location of the HST/WFPC2
detectors is shown on the map of detected RRab stars in Fig.~5. About
3.9\% of the whole cluster area falls in the gaps between the OGLE CCDs.

\begin{figure}[htb]
\includegraphics[width=12.5cm]{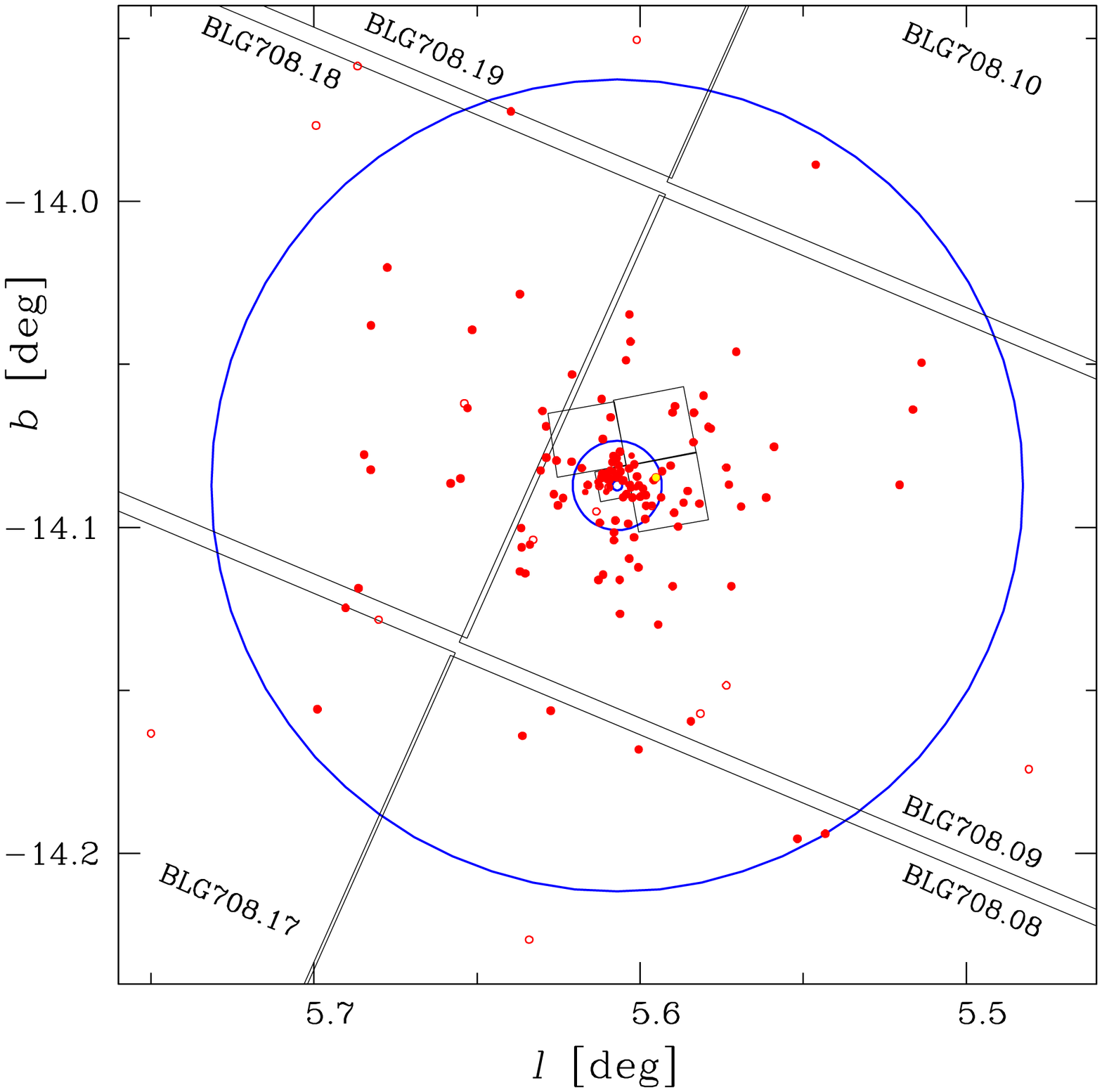}
\FigCap{RRab stars in the field of the globular cluster M54. Blue
circles represent the characteristic cluster radii: core
($0\zdot\arcm09$), half-light ($0\zdot\arcm82$), and tidal
($7\zdot\arcm5$). Long black lines delimit the OGLE-IV chips. Central
black squares show the area observed with the HST/WFPC2 camera. RRab
stars belonging to the cluster are marked with filled red circles, while
field stars are marked with empty circles. The unusually long-period
RRab variable V92, located approximately at half-light radius, is marked
with a yellow circle. All RRab stars located outside the tidal radius
were classified as field objects.}
\end{figure}

The search for variable stars was done in three stages. In the
beginning, we searched for any kind of variable objects among stars
detected with the standard OGLE reduction pipeline which is based
on the image subtraction technique (\cf Udalski \etal 2015). We limited
ourself to objects detected within the tidal radius of the cluster. All
light curves were then subject to the period search. We used OGLE
standard periodogram procedure -- {\sc Fnpeaks} Fourier analysis
software\footnote{http://helas.astro.uni.wroc.pl/deliverables.php?lang=en\&active=fnpeaks},
providing the most prominent periods and their signal-to-noise.

We made a visual inspection of $I$-band light curves of 4711 detections
with a significant periodicity signal (S/N$>4$) selected from a total
number of 53~297 point sources. After rejection of some false
identifications due to blending and crowding, particularly strong in the
inner part of the cluster, 192 candidate variable stars were left. Then,
we identified the OGLE counterparts with variable objects in the updated
Clement \etal (2001) list. The identification was generally based on
coordinates, but also on observed brightness and variability
time-scales. We found 24 objects to be non-variable at all and eight
variable stars misclassified. Thanks to the long-term OGLE observations we
could obtain more accurate periods for pulsating variable stars and eclipsing
binary stars. Relative period uncertainties for these types of periodic
variable stars are usually between $10^{-6}$ and $10^{-5}$, as returned by
the TATRY code (Schwarzenberg-Czerny 1996). In the case of semi-regular
variable red giants, errors of estimated periods are of 1~d or larger.

We also verified, yet unconfirmed, candidates for RR Lyr stars found by
Montiel and Mighell (2010) in the archival HST/WFPC2 images of the
central area of M54 (program GO~6701). Ten of the stars located in
moderately crowded regions were easily cross-matched with the OGLE
detections. In the case of the remaining objects, we encountered a
problem of the lack of any detections from the standard OGLE image
reduction pipeline. It was a result of generating $11 \times 11$ pixel
masks in the location of bright stars with over 60~000 counts per pixel
to secure linear transformation to magnitudes. We found that the
brightest stars in the center of M54 did not in fact reach the
saturation limit of 65~536 counts in our frames and crossed the
60~000 limit in just single pixels.

In this situation, we decided to perform a second stage of variable
stars search by reducing in a different way the data for a central 3.0
arcmin$^2$ area. First, we lowered the mask size to only one pixel.
We chose eleven difference images with the lowest seeing. Then, after
taking the absolute value of all residuals we stacked the images
together creating a variability image. In such a combined image, the
signal from variable sources was expected to be much stronger. Searches
for positive detections resulted in finding of 2730 candidate variable
sources. In the position of each source, we extracted $I$-band
time-series photometry. After visual inspection of the new light curves,
rejection of false positives, and cross-matching with the updated
Clement \etal (2001) list, we were left with 85 variable objects.

Positions of only three objects coincided with the positions of
unconfirmed candidate RR Lyr stars from HST (Montiel and Mighell 2010).
Among other sources we found 14 new RR Lyr type stars and two new
Type II Cepheids. All variable stars were calibrated from the instrumental
to the standard magnitudes using transformation relations and
coefficients given in Udalski \etal (2015). In the case of several
new variable stars, we assumed the average color of other variable stars of the
same type due to severe crowding and lack of $V$-band information.

Finally, as a third stage, we decided to extract OGLE $I$-band
photometry in the exact locations of all 50 HST candidate RR Lyr stars.
Prior to that we reduced the archival WFPC2 images using the {\sc
HSTphot} package (Dolphin 2000ab). We performed some initial image
processing steps like masking bad pixels and cosmic-ray rejection.
Profile photometry was obtained using a library of model point-spread
functions and transformed to the standard $V$ and $I$ bands using the
utility available in the {\sc HSTphot} package. (X,Y) positions of
candidates on the HST images were transformed to the (X,Y) grid of OGLE
reference images to allow photometry at correct position on the OGLE
difference images. If the resulting light curve contained periodic
signal in the suitable period range and the shape resembled that of RR
Lyrae stars the detection was considered as positive.

We confirm the presence of another seven RR Lyr stars and assess
their periods based on the OGLE data. It should
be stressed here that neither of the new RR Lyrae could be clearly
resolved on the OGLE reference image due to severe crowding in the central
part of the cluster in spite of its good resolution of $\approx 1\arcs$.
Nevertheless, the difference signal was sound in all these cases.

Unfortunately, we are not able to detect any variability signal at the
positions of the remaining 16 candidates (50 minus 14
cross-identified with the Clement's \etal list minus 20 detected during
our search in stages 1--3) indicated by Montiel and Mighell (2010) in
the HST/WFPC2 images. To calibrate the confirmed RR Lyr variable stars we
scaled and shifted the phased OGLE $I$-band light curves to match the
brightness returned by {\sc HSTphot}. Due to severe crowing in the
central part of the cluster the accuracy of zero points for these stars
can be as bad as 0.1~mag.

Summarizing, our search for variable stars in the M54 cluster field
led to the discovery of 83 new variable stars.  We also cross-identified
and verified 205 objects from the Clement's \etal (2001) list of M54
variable objects containing 211 stars. We also note that there is no
OGLE photometry for variable stars V14, V56, V65, V66, V108, and V157. These
stars fell in the gaps between the OGLE CCDs.

\section{Detected Variable Stars}

The final list of variable stars in the field of the globular cluster
M54, time-series $V$- and $I$-band photometry and finding charts are
available to the astronomical community from the OGLE Internet Archive:
\begin{center}
{http://ogle.astrouw.edu.pl\\
ftp://ftp.astrouw.edu.pl/ogle/ogle4/OCVS/M54/\\}
\end{center}

Originally, we planed to continue Clement's \etal (2001) scheme of
numbering the variable stars by assigning the name 'V212' (the first
available new variable) to our first variable and so on. In the first
publicly available version of our paper such numbering was presented.
However, because the results of independent search for variable stars in
the central regions of M54 conducted by the MiNDSTEp consortium were
published practically simultaneously with our results (Figuera Jaimes \etal
2016) and they introduced similar to our numbering scheme of their
variable objects we jointly undertook efforts to synchronize the naming
convention to avoid ambiguities and confusion for the future researchers
working on the globular cluster M54.

We decided to rename the original OGLE names and assign names V212 to
V291 for objects presented in the MiNDSTEp paper (Table~3, Figuera Jaimes \etal
2016). The remaining OGLE new variable stars -- not present on the MiNDSTEp
list -- follow as variables V292--V347. However, one has to remember
that 46 out of 76 MiNDSTEp objects (V212--V291) were also found in this
study and these are independent discoveries. 

Figuera Jaimes \etal (2016) used Electron-Multiplying CCD detector and
technique allowing obtaining much better image resolution, thus they
were able to search for variable stars in the most dense part of M54,
practically not accessible for the standard OGLE observing
setup. Most of the OGLE missing variable stars are located in the very center
of the cluster.

For 20 RR Lyr stars from the HST candidate list (Montiel and Mighell
2010) confirmed as genuine pulsating stars in this OGLE study, 16 have
been also listed by MiNDSTEp. We assign IDs from V292 to V295 for the
remaining 4. Other newly discovered variable stars have numbers from V296
to V347 and are sorted with increasing right ascension. In the file
listing the variable stars ({\sf M54variables.dat} in the OGLE Internet
Archive), we provide for each object information on coordinates,
distance from the center of the cluster, variability type, detected
period, $V$- and $I$-band brightnesses, and $I$-band amplitude, if
available. Remarks on some stars are given in the last column.

We confirm that objects already marked by Clement \etal (2001) as
constant stars do not show any brightness variations, \ie positions:
V20, V21, V22, V24, V26, V27, V53, V72, and V100. Surprisingly, 28
variable stars (mostly classified as RR Lyr-type stars) do not vary at all.
These are objects: V73, V79, V81, V86, V91, V107, V155, V166, V167,
V169, V170, V175, V189, V195, V196, V197, V198, V199, V200, V201,
V202, V203, V204, V205, V207, V208, V209, and V211. Among objects
from V190 to V211 reported by Sollima \etal (2010),
fifteen are in fact very bright constant stars
located at a distance of a few $r_{\rm h}$ from the cluster center.
Several objects from V196 to V210 had very suspicious periods that
closely grouped either around a value of 0.35~d or 0.50~d. 
Bright objects (with $I \approx 13$ mag) named V190, V191, V206, and
V210 are not RR Lyr-type stars but variable red giants. The same
conclusion refers to variable V71.

We confirm that Soszy\'nski \etal (2014) correctly classified variable
V12 as a RRd star. They also correctly identified variable stars V172, V184,
and V192 as RRab stars, not RRc stars as it had been thought before.
These three variable stars have light curves typical for RRab stars and
pulsation periods longer than 0.6~d. For a comparison, RRc type stars
have pulsation periods in the range between 0.20~d and 0.54~d.

Among RR Lyr stars discovered by Soszy\'nski \etal (2014), 19 variable stars
are located within the M54 tidal radius and are not included in the
updated list by Clement \etal (2001). These are stars:\\
V227=OGLE-BLG-RRLYR-37582, V228=OGLE-BLG-RRLYR-37575,\\
V229=OGLE-BLG-RRLYR-37597, V233=OGLE-BLG-RRLYR-37591,\\
V236=OGLE-BLG-RRLYR-37585, V237=OGLE-BLG-RRLYR-37570,\\
V240=OGLE-BLG-RRLYR-37576, V244=OGLE-BLG-RRLYR-37581,\\
V246=OGLE-BLG-RRLYR-37573, V250=OGLE-BLG-RRLYR-37590,\\
V251=OGLE-BLG-RRLYR-37579, V252=OGLE-BLG-RRLYR-37593,\\
V253=OGLE-BLG-RRLYR-37586, V295=OGLE-BLG-RRLYR-37578,\\
V296=OGLE-BLG-RRLYR-37500, V310=OGLE-BLG-RRLYR-37565,\\
V321=OGLE-BLG-RRLYR-37594, V330=OGLE-BLG-RRLYR-37621\\
and V338=OGLE-BLG-RRLYR-37662. Variable V321 is the second known RRd
type star in M54.

Among the 50 HST candidates for RR Lyr variable stars proposed by Montiel and
Mighell (2010), 14 stars were indeed known before as noticed by
C.~Clement: V127=VC2, V162=VC11, V163=VC12, V95=VC13, V164=VC14,
V142=VC15, V129=VC17, V179=VC18, V181=VC28, V160=VC34, V46=VC44,
V148=VC45, V192=VC46, and V76=VC47. With the OGLE photometry we confirm
that the following 20 variable stars are RR Lyr stars: V213=VC40, V214=VC39,
V215=VC25, V220=VC4, V223=VC10, V231=VC26, V233=VC30, V234=VC7,
V236=VC36, V237=VC31, V241=VC1, V250=VC23, V255=VC5, V284=VC32,
V285=VC33, V291=VC27, V292=VC16, V293=VC22, V294=VC24, and V295=VC38.
Five of the stars, namely V213, V215, V291, V292, and V294, turned out
to be of RRc type, while the remaining objects are of RRab type. Sixteen
HST objects could not be confirmed with the OGLE data. It is worth
noting that MiNDSTEp study (Figuera Jaimes \etal 2016) based on better
resolution images and reaching more central parts of M54 confirmed
variability of additional six HST objects: V212=VC35, V216=VC9,
V221=VC8, V224=VC20, V225=VC37, and V232=VC41.

We also confirm classification of seven additional RR Lyr stars found
by the MiNDSTEp consortium (Figuera Jaimes \etal 2016) and overlooked
during our search: V218, V222, V235, V238, V243, V248, and V282.

Our search for new variable stars has led to the discovery of the following
83 objects: 26 RR Lyr-type stars, two Type II Cepheids, nine eclipsing
binaries, two ellipsoidal binaries, 22 semi- or irregular variable red
giants, and 22 unclassified variable stars. Among the RR Lyr variable stars, there
are 18 RRab and eight RRc stars. $I$-band light curves of these new RR
Lyr variable stars are shown in Figs.~6 and 7. In Fig.~8, we present light
curves of all four Type II Cepheids. In addition to the two previously
known BL Her-type stars (V1 and V2), we have discovered one more BL
Her-type star (V230) and one W Vir-type star (V256). Among eclipsing
binaries, we have found four EW systems and five EB systems. Light
curves of eclipsing binary stars and semi-regular variable red giants
are shown in Figs.~9 and 10, respectively. In the case of variable stars of
unknown type, more data is required to properly classify them.

\begin{figure}[htb]
\includegraphics[width=12.5cm]{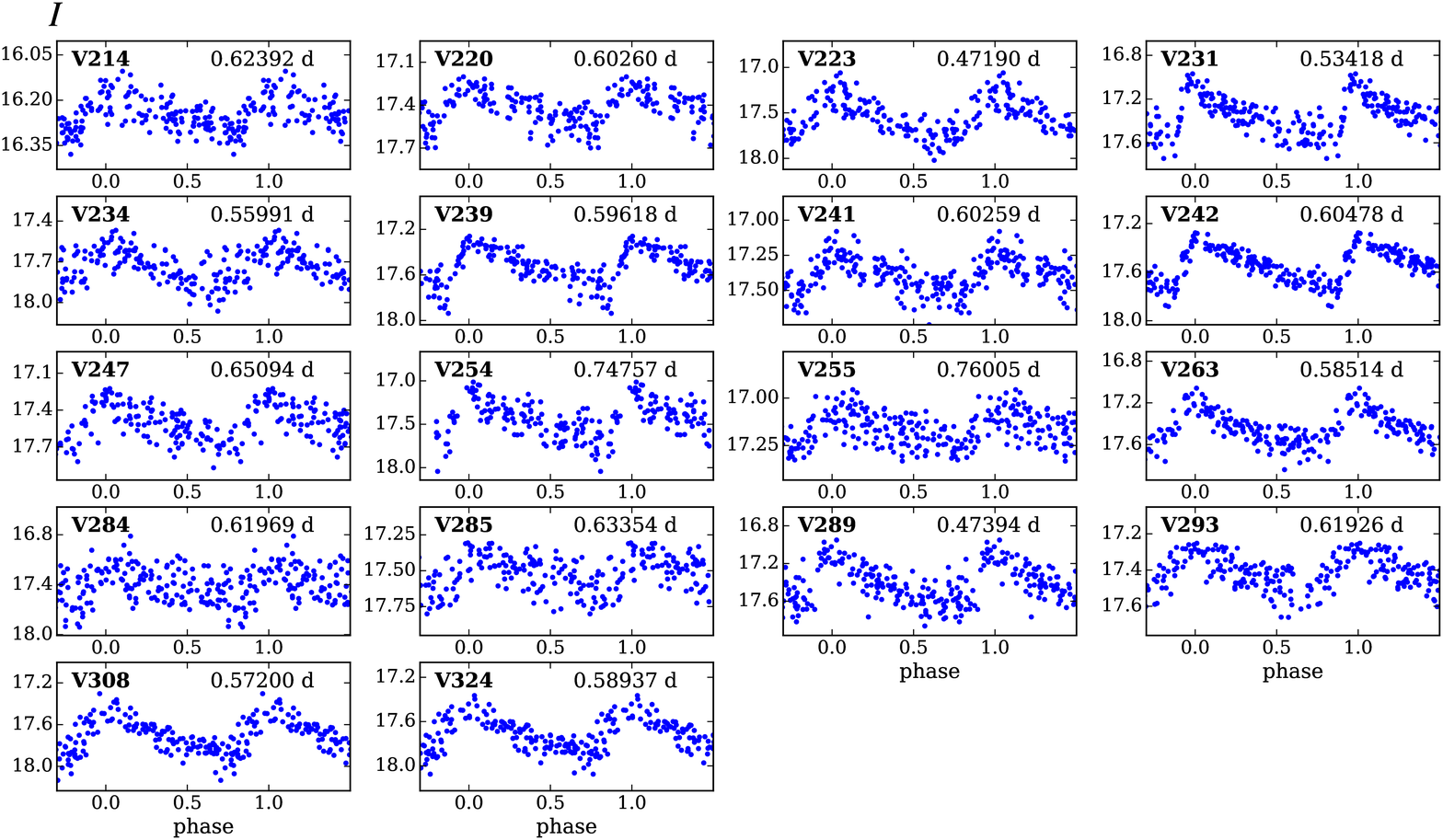}
\FigCap{Phased OGLE $I$-band light curves of new RRab variable stars from M54.}
\end{figure}

\begin{figure}[htb]
\includegraphics[width=12.5cm]{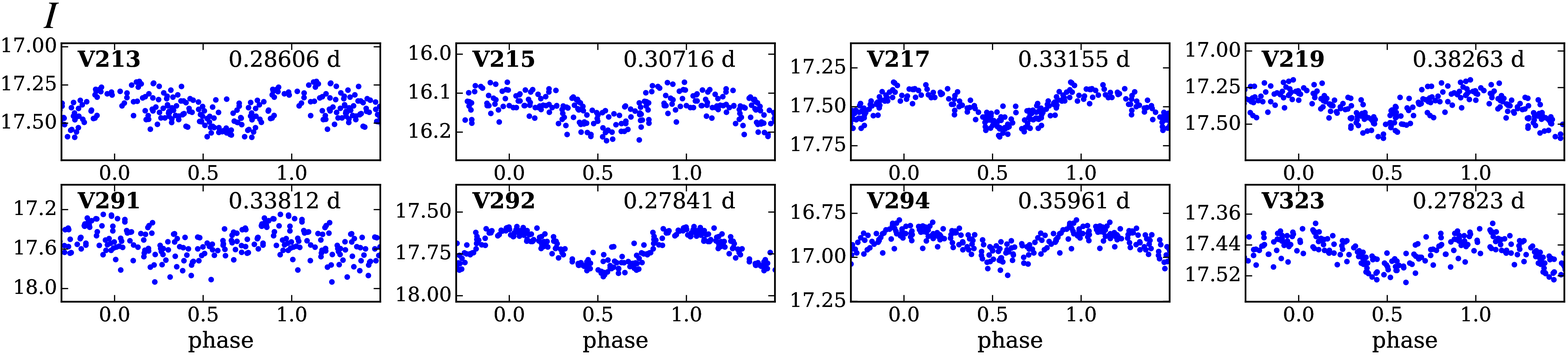}
\FigCap{Phased OGLE $I$-band light curves of new RRc variable stars from M54.}
\end{figure}

\begin{figure}[htb]
\begin{center}
\includegraphics[width=10.0cm]{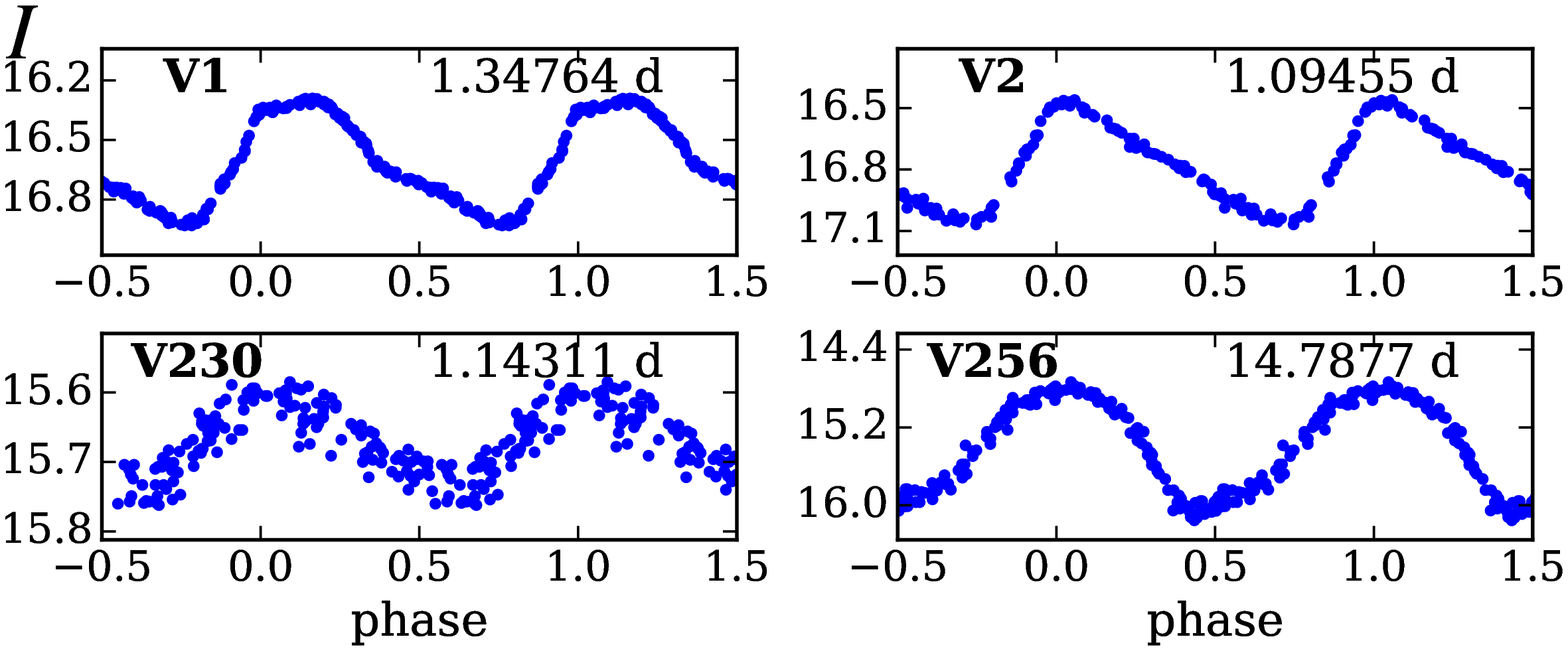}
\end{center}
\FigCap{Phased $I$-band light curves of all known Type II Cepheids in
M54. Two bottom objects are newly discovered pulsators.}
\end{figure}

\begin{figure}[htb]
\includegraphics[width=12.5cm]{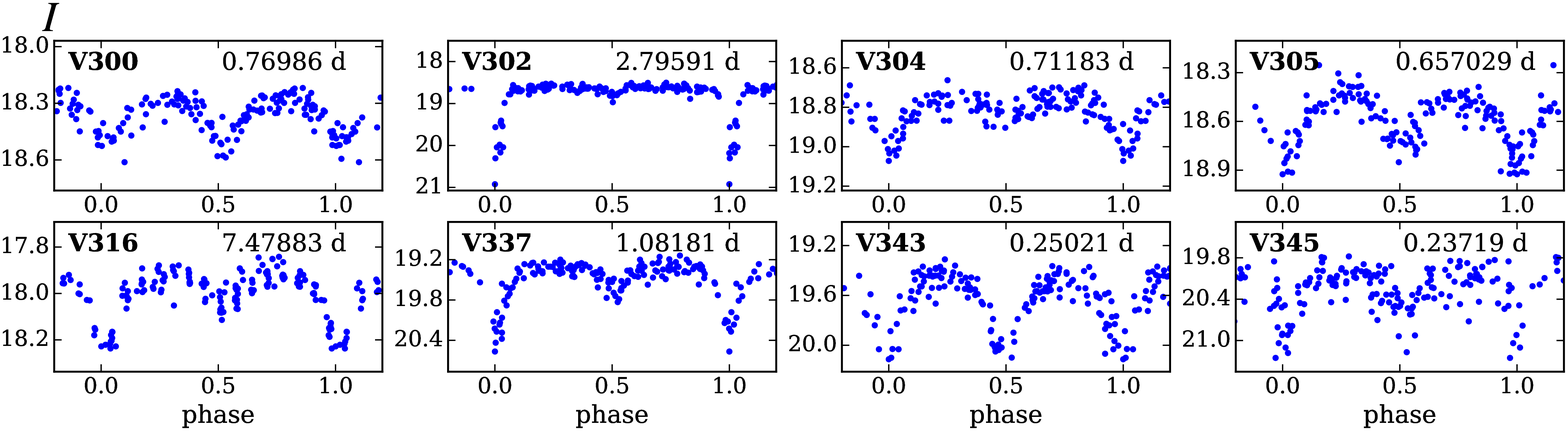}
\FigCap{Phased $I$-band light curves of selected eclipsing binary stars
in the field of M54.}
\end{figure}

\begin{figure}[htb]
\begin{center}
\includegraphics[width=12.5cm]{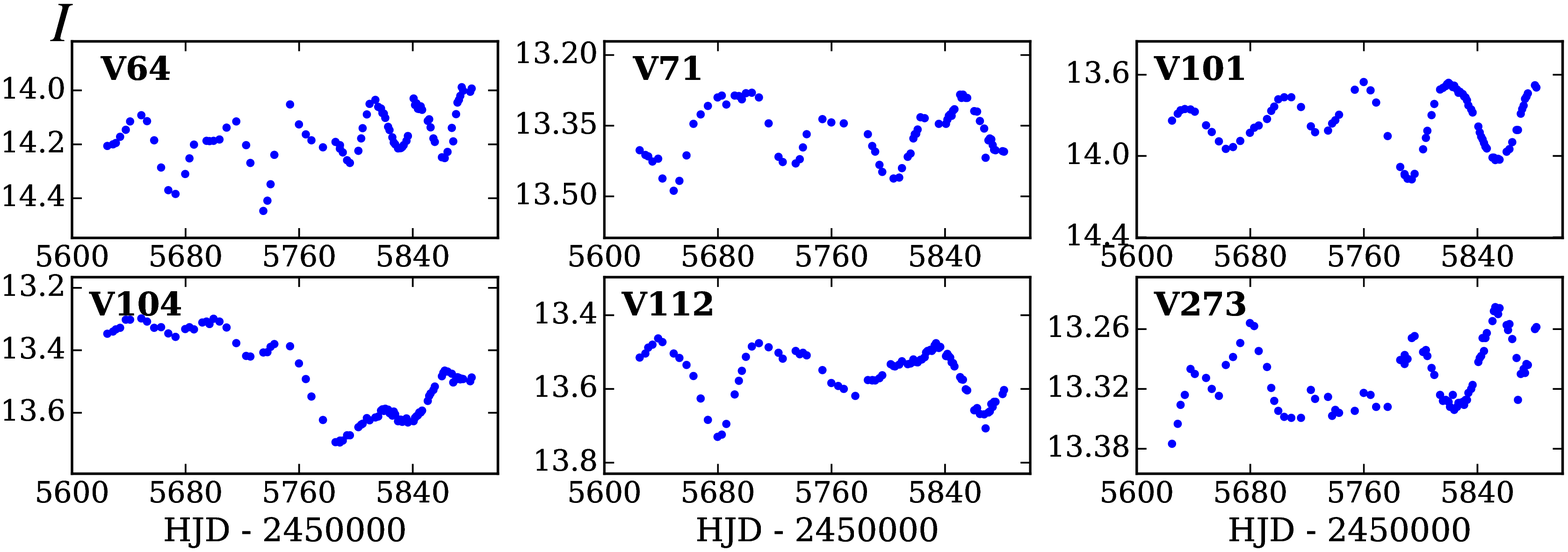}
\end{center}
\FigCap{$I$-band light curves of selected semi-regular variable red
giants from M54 observed during 2011 season.}
\end{figure}

In the upper panel of Fig.~11, we present color-magnitude diagram (CMD)
for the globular cluster M54 with positions of variable stars with
measured brightness in the $V$ and $I$ bands. The background is formed
of all stars detected within the tidal radius of the cluster with the
OGLE standard pipeline. We only show variable stars located between the
half-light and tidal radii. For comparison, in the lower panel of Fig.
11, we present CMD for a nearby off-cluster field containing mostly
stars from the Sgr dSph galaxy. Common features are present in both
diagrams. Foreground main sequence stars from the Galactic halo are seen
in the area between $V-I=0.8$ mag and $V-I=0.9$ mag for $I<19.5$~mag.
Red giants from the Galactic halo and Sgr dSph galaxy are spread along a
wide arc from ($V-I$,$I$)=(1.0,19.5) toward upper right corner. Sgr dSph
red clump stars can be found around ($V-I$,$I$)=(1.2,17.1). In the
diagram for M54, there are additional features like prominent red giant
branch with the red clump at ($V-I$,$I$)=(1.15,17.0) and horizontal
branch occupied by RR Lyr stars with $0.45<V-I<0.85$ mag. Note that the
M54 red clump does not overlap with the Sgr dwarf red clump. The
location of the M54 red clump slightly to the blue of the Sgr dSph red
clump indicates that the cluster is more metal poor than the remaining
body of the dwarf galaxy.

\begin{figure}[htb]
\includegraphics[width=12.5cm]{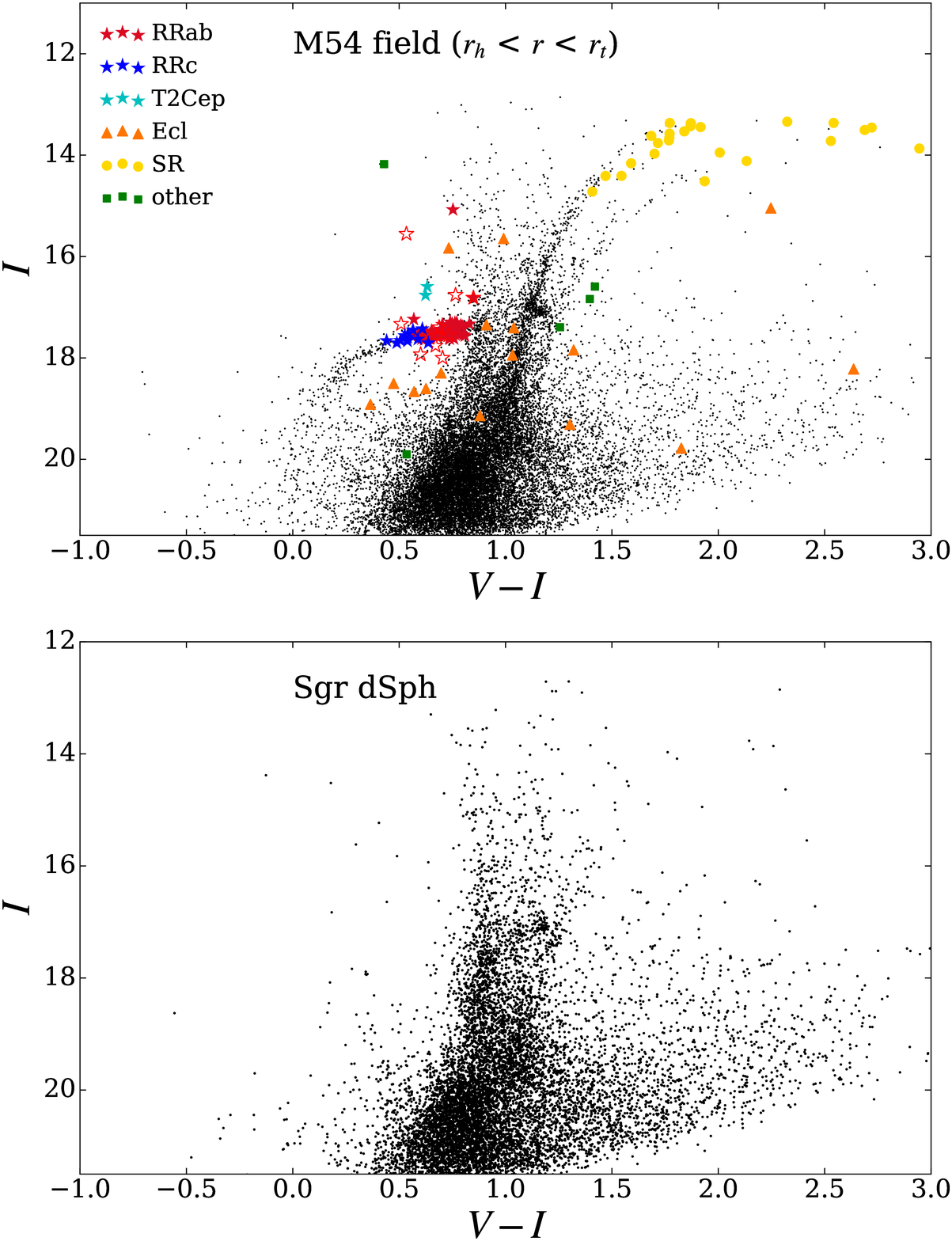}
\FigCap{$I$ vs. $V-I$ diagrams for the M54 cluster field (\textit{upper panel})
and an off-cluster area (lower panel). The \textit{upper panel} shows stars
located between the half-light radius ($r_{\rm h} = 0\zdot\arcm82$) and
tidal radius ($r_{\rm t} = 7\zdot\arcm5$) of the cluster. Variable stars
of different types are marked with different symbols. Open red asterisks
denote non-member RRab stars. The \textit{lower panel} presents stars from the
southern half of the OGLE-IV chip BLG708.09. Note the prominent features
of M54 such as the red giant branch and horizontal branch. The M54 red
clump can be found to the blue of the Sgr dSph red clump. This indicates
a lower metallicity of the cluster in comparison to the parent dwarf
galaxy.}
\end{figure}

Based on measured brightness we confirm that the following RRab stars
located between the half-light and tidal radii are not members of the
globular cluster: V55, V65, V93, V119, and V121. We also confirm that
RRab variable V92 is a field object despite its location at only
0.74~$r_{\rm h}$ from the center of M54. This star with a pulsation
period 0.4847065(10)~d is brighter than the M54 horizontal branch by
about 0.35 mag. According to Clement \etal (2001) V92 is blended.
Indeed, it has a neighbor of a similar brightness, but located about
$1\zdot\arcs3$ (5 pix) away in OGLE images. At such distance the
neighbor star cannot affect much the measured brightness of the
variable.

The total number of RR Lyr stars considered as cluster members is 163.
We assume that all variable stars within $r_h=0\zdot\arcm82$ but the
mentioned V92 belong to M54. Based on brightness information given in
Clement \etal (2001) we assume that RR Lyr variable stars V14, V30, V56, V75,
which fell in the gaps between the OGLE chips, belong to the cluster.
Among the 163 members there are 133 RRab stars, 28 RRc stars, and 2 RRd
stars. Fundamental-mode RR Lyr stars being non-members of M54 are marked
with open circles on the map in Fig.~5.

The membership status of V184 requires a few words of explanation. This
variable is located at 0.86~$r_{\rm h}$ from the center. Its pulsation
period is 0.959234(8)~d. It is much longer than the average for the
cluster (0.599~d). V184 is brighter than the cluster horizontal branch
by about 0.50~mag. We suppose that this object is an M54 RR Lyr star
probably leaving the horizontal branch toward the asymptotic giant
branch and crossing the instability strip. In future, it would be worth
to verify whether its period increases due to its evolution redward
across the instability strip. We mark this object with a filled yellow
circle in Fig.~5.

Positions of Type II Cepheids V1 and V2 in the cluster CMD (Fig.~11)
confirm that these stars are members of the cluster. The two newly
discovered Type II Cepheids, BL Her-type variable V230 and W Vir-type
variable V256, are located well within the half-light radius of M54, at
0.18 $r_{\rm h}$ and 0.12 $r_{\rm h}$, respectively. Their estimated
mean $I$-band brightness is 15.67 mag and 15.29~mag, respectively. These
facts clearly indicate that the two Cepheids also belong to the cluster.

Based on the position in the CMD, we conclude that probably all
semi-regular variable red giants detected within the tidal radius of M54
are likely cluster members. Observed binary stars and other unclassified
variable stars are likely foreground objects.

\section{Distance to M54}

Based on 71 RRab stars considered as members of M54 and located
between its half-light and tidal radii to minimize the effect
of crowding, and for which we have information on brightness in $V$ and $I$,
we estimate the distance to this globular cluster. Using the presented
sequence of equations (1--9) we find $d_{\rm M54} = 26.7 \pm 0.03_{\rm stat}$~kpc.
This is the median value of individual distances to these stars.
The statistical uncertainty is calculated using error propagation
formula applied to Eq.~8 assuming a mean accuracy of OGLE photometric
measurements $\sigma_I=0.02$~mag and a very small mean random error
of $\phi_{31}$ of 0.018. Uncertainties of the pulsation periods
are negligible. The final value of the statistical
uncertainty was divided by square root of the number of used stars.
The largest contributions to the total systematic uncertainty are
errors of the zero point of the period-luminosity-metallicity relation
(Eq.~3), the conversion for metallicity (Eq.~2), and uncertainty
of the $E(B-V)$ map (Chen \etal 1999). The systematic error on distance
modulus can be as large as 0.1~mag (Pietrukowicz \etal 2015),
which at the distance of M54 translates to 1.3~kpc.

\section{Multiple Old Populations in M54 and the Sgr dSph Stream}

We use the detected RR Lyr stars to look for old stellar populations,
both in the cluster and main body of the dwarf galaxy. In Fig.~12, we
compare period distributions for all RRab stars from M54 and the
Sgr stream. The mean period for all known cluster members is 0.599~d,
not excluding the long-period variable V184. For a comparison, the mean period
for the Sgr dwarf RRab stars is slightly lower, 0.579~d. We applied the
Kolmogorov-Smirnov statistics to compare the distributions. We find the
p-value below 1\% which indicates that the distributions are almost
identical. 

\begin{figure}[htb]
\includegraphics[width=12.5cm]{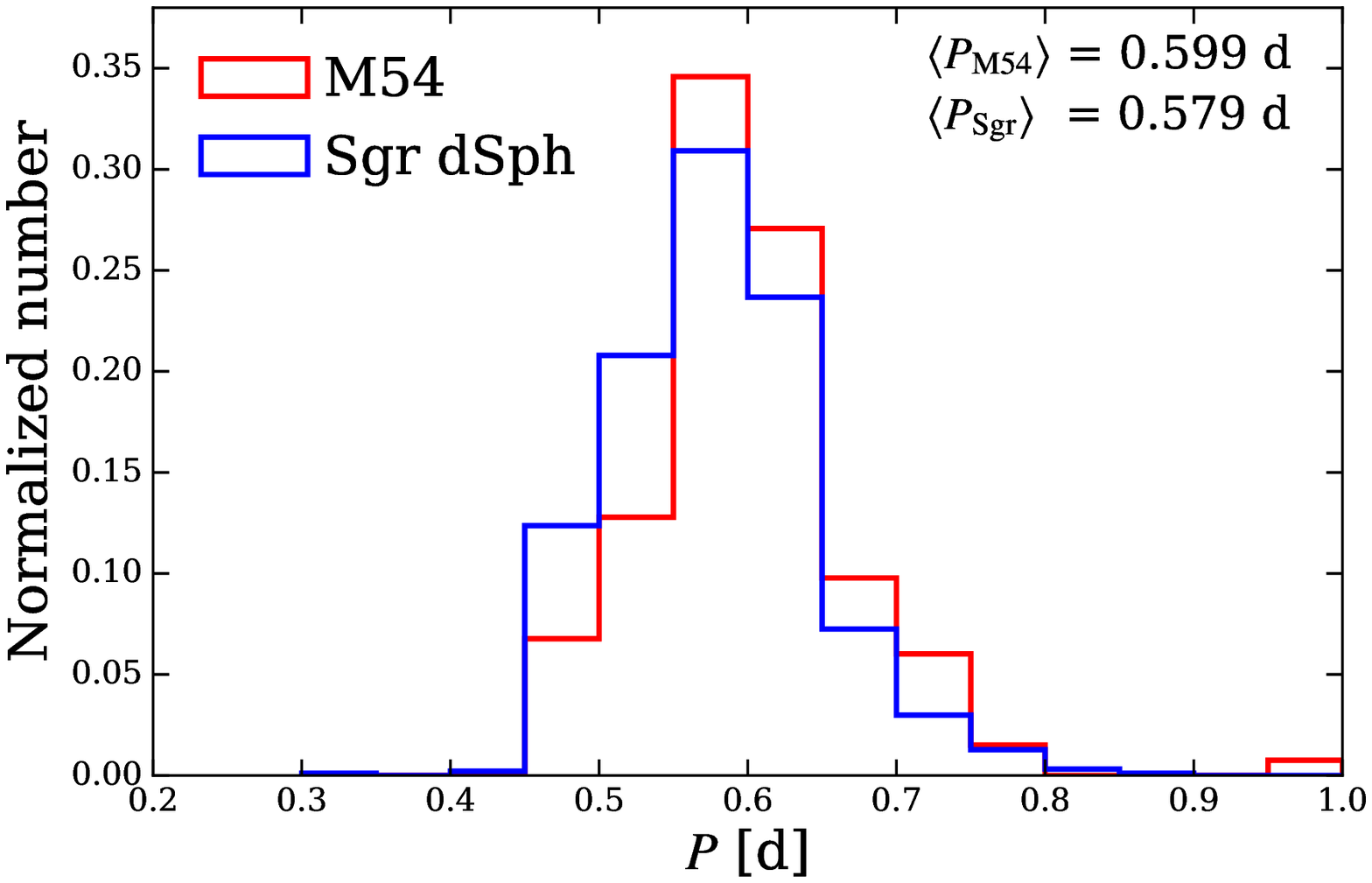}
\FigCap{Comparison of period distributions for RRab stars from the
globular cluster M54 and Sgr dSph galaxy.}
\end{figure}

In Fig. 13, we plot period--amplitude (or Bailey) diagram for cluster
and Sgr dSph galaxy fundamental-mode RR Lyr stars. To minimize the
effect of crowding on the measured $I$-band amplitude we only show
variable stars located between the half-light and tidal radii of the cluster.
In both systems, the cluster and dwarf galaxy, RRab stars form two
sequences, which correspond to Oosterhoff groups OoI and OoII.
Despite a low number of M54 RRab variable stars representing
the OoII group (only 10), it is very unlikely that these objects are contaminants
from the Sgr dwarf. This number of objects gives an average density
of 0.057 arcmin$^{-2}$ for the cluster. For a comparison, in the
whole off-cluster area within the OGLE-IV field BLG708 covering 
$\approx1.35$~deg$^2$ there is about 30 such variable stars (\ie objects
above the broader sequence in Fig.~13), which gives 0.006 arcmin$^{-2}$
or roughly ten times lower density for the Sgr dwarf.
Approximate ratio of detected OoII stars to OoI stars in the cluster
and dwarf galaxy is very similar: 14\% and 16\%, respectively.
This is roughly twice smaller to what was found in the Sgr dSph
stream far away from the core (\eg Zinn \etal 2014).
The observed bimodal Oostherhoff distribution in the dwarf galaxy
and globular cluster indicates the presence of two metal-poor
populations which differ by about 0.2~dex (Fig.~14): OoI of
[Fe/H]$_{\rm J95}\approx-1.2$~dex and OoII of [Fe/H]$_{\rm J95}\approx-1.4$~dex.
No bimodal distribution in the metal-poor regime has been found
in spectroscopic surveys of red giants from M54 and Sgr nucleus
(Bellazzini \etal 2008, Carretta \etal 2010), but they did not
include RR Lyr stars at all. Similar ratios of OoII to OoI variable stars
in the cluster and in the dwarf galaxy suggest the same formation
scenario of both systems.

\begin{figure}[htb]
\includegraphics[width=12.5cm]{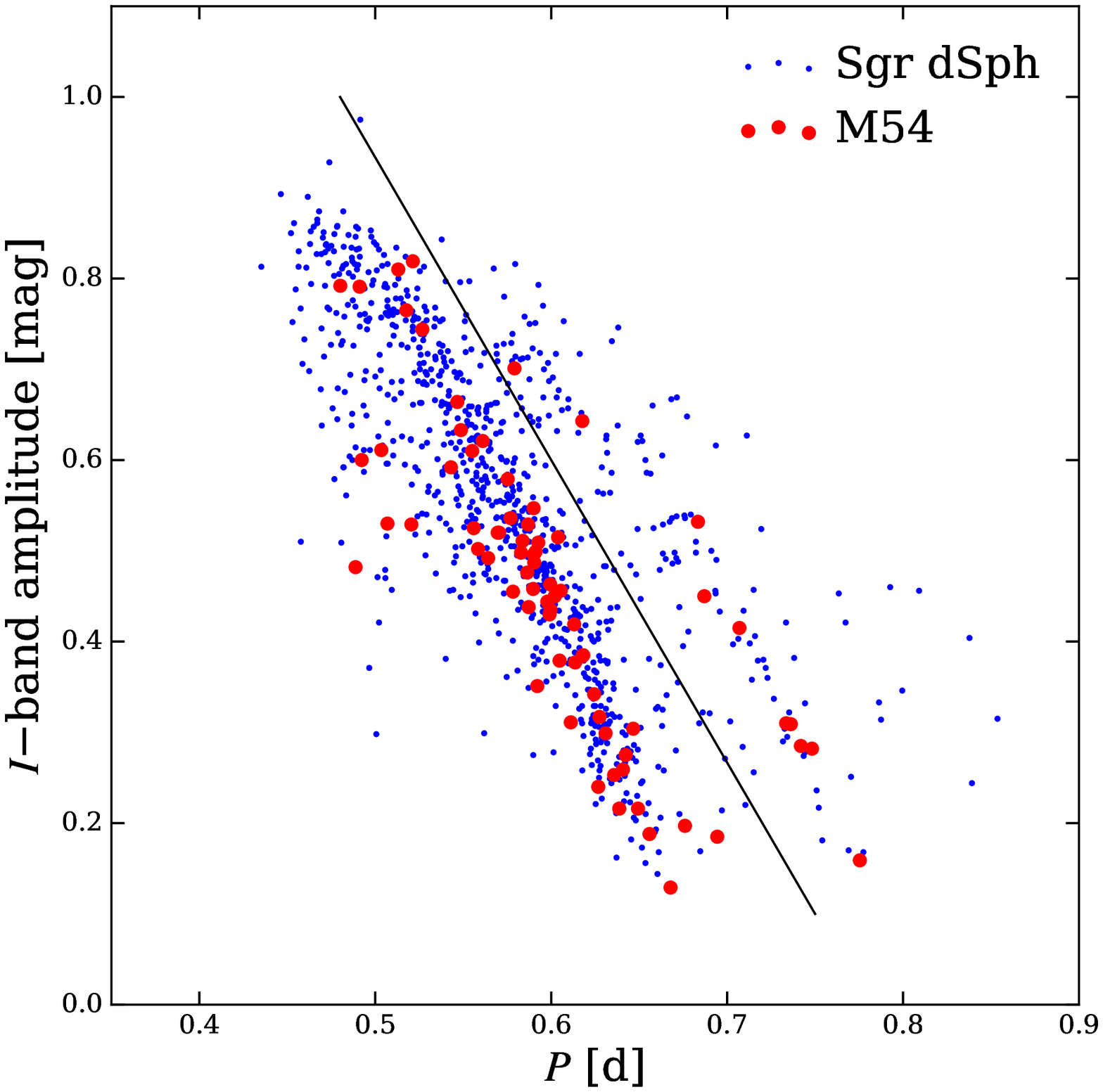}
\FigCap{Bailey diagram for RRab stars from M54 (red points) and main
body of the Sgr dSph galaxy (blue points). Straight line roughly
divides the variable stars into two Oosterhoff groups: OoI group at shorter
periods and OoII at longer periods. Two sequences indicate the
presence of two old populations both in the cluster and Sgr dwarf.}
\end{figure}

\begin{figure}[htb]
\includegraphics[width=12.5cm]{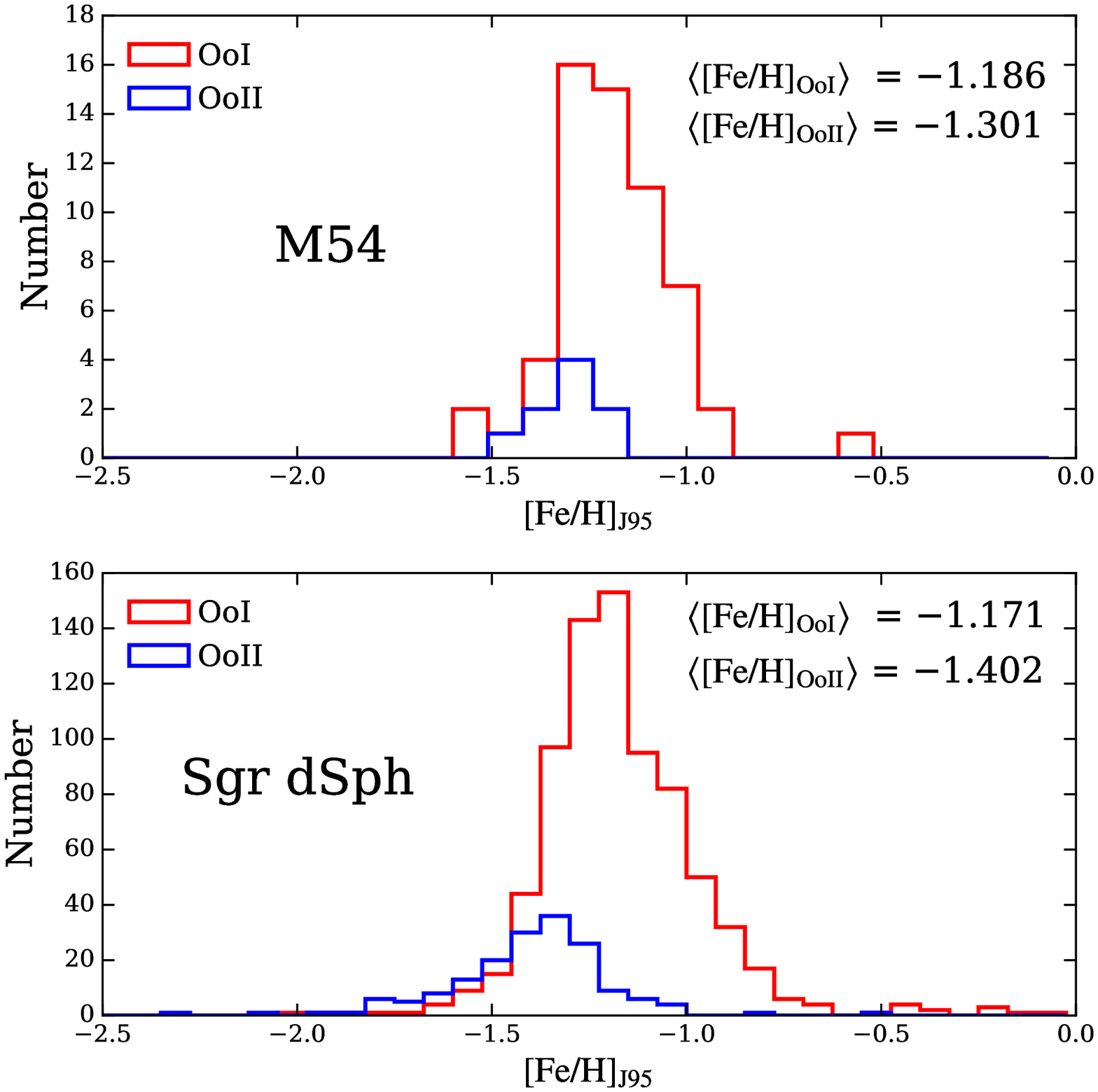}
\FigCap{Comparison of metallicity distributions for RRab stars
representing two Oosterhoff groups from M54 (\textit{upper panel}) and
the Sgr dSph galaxy (\textit{lower panel}). The metallicities are
on the Jurcsik (1995) scale.}
\end{figure}

\section{Conclusions}

We presented the 3D picture of the Sgr dSph galaxy observed roughly in
the background of the Galactic bulge based on RRab stars. We estimated
the line-of-sight thickness of the central section of the Sgr dSph
stream to be FWHM$_{\rm cen}=2.42$~kpc. We also presented the results of
our comprehensive search for variable stars in the field of the globular
cluster M54 residing in the core of the Sagittarius Dwarf Spheroidal
galaxy. Our aim was to verify and expand the existing list of variable
stars in M54. We have confirmed the presence of 20 RR Lyr-type stars
indicated in the archival HST/WFPC2 images by Montiel and Mighell
(2010). We have discovered 83 new variable stars, 26 of which are RR Lyr stars.
We have also reviewed the previously suggested variable stars, showing that 24
variable stars, mostly classified as of RR Lyr type, are in fact non-variable.
Using RRab pulsators, we have estimated the distance to the M54 cluster
obtaining $d_{\rm M54} = 26.7 \pm 0.03_{\rm stat} \pm 1.3_{\rm sys}$ kpc.
We have confirmed the presence of two old populations, both in the
cluster and dwarf galaxy.

\Acknow{We would like to thank Profs. M. Kubiak and G. Pietrzy\'nski,
former members of the OGLE team, for their contribution to the
collection of the OGLE photometric data over the past years. We also
thank Prof.~V.~Belokurov for discussion and suggestions and Drs.
R.~Figuera Jaimes and D.~Bramich for informing us about MiNDSTEp consortium
study of the M54 center and joint work on unambigous numbering of newly
discovered variable stars. We acknowledge the referee for very constructive
comments.

The OGLE project has received funding from the National Science Centre,
Poland, grant MAESTRO 2014/14/A/ST9/00121 to A.U. This work has been
also supported by the Polish Ministry of Sciences and Higher Education
grants No. IP2012 005672 under the Iuventus Plus program to P.P. and No.
IdP2012 000162 under the Ideas Plus program to I.S.}

\end{document}